\documentclass[12pt]{iopart}

\usepackage{graphicx}
\usepackage{dcolumn}
\usepackage{bm}
\usepackage{amssymb}
\usepackage{multirow}
\usepackage{iopams}
\usepackage{color}

\begin{document}

\title{Correlations and pair emission in the escape dynamics of ions from one-dimensional traps}

\date{\today}

\author{C. Petri$^1$, S. Meyer$^2$, F. Lenz$^1$, and P. Schmelcher$^{1}$}
\address{$^1$ Zentrum f\"ur optische Quantentechnologien, Universit\"at Hamburg, Luruper Chaussee 149, 22761 Hamburg, Germany}
\address{$^2$ Physikalisches Institut,  Universit\"at Heidelberg, Philosophenweg 12, 69120 Heidelberg, Germany}
\eads{\mailto{cpetri@physnet.uni-hamburg.de}, \mailto{flenz@physnet.uni-hamburg.de}, \mailto{pschmelc@physnet.uni-hamburg.de}}

\begin{abstract}
We explore the non-equilibrium escape dynamics of long-range interacting ions in one-dimensional traps. The phase space of the few ion setup and its impact on the escape properties are studied. As a main result we show that an instantaneous reduction of the trap's potential depth leads to the synchronized emission of a sequence of ion pairs if the initial configurations are close to the crystalline ionic configuration. The corresponding time-intervals of the consecutive pair emission as well as the number of emitted pairs can be tuned by changing the final trap depth. Correlations between the escape times and kinetic energies of the ions are observed and analyzed.
\end{abstract}

\maketitle

\section{Introduction}
The controlled manipulation of ions in traps has triggered several areas in physics. Especially the impact on the development of high precision measurements in spectroscopy \cite{Thomp:1993} and on concepts for implementing quantum simulations (see \cite{Wineland:2003,Blatt:2008,Wunder:2009} and references therein) has been substantial. An interesting effect observed in ion-trap resonators is the synchronization of ionic motion \cite{Zajfman:2001} with applications in mass spectrometry. Furthermore, the study of the stability and the loss processes of ions shows that in linear electrostatic traps up to 30 \% of trapping efficiency is reachable \cite{Zajfman:2002}. By applying the $\delta$-kick method a compression of the velocity spread of stored ions in these traps has been demonstrated \cite{Zajfman:2003}, which can be exploited for cooling. Another widely-used ion trap is the so-called Paul trap \cite{Paul:1990} which has been used to observe and monitor single ions \cite{Dehmelt:1978,Dehmelt:1978b,Dehmelt:1980} or to study the stability of large Coulomb ion crystals \cite{Schifffer:1998,Drewsen:2002,Drewsen:2003}. Originally, the first experimental observation of a phase transition from a cloud of ions to a crystalline structure has been reported in Ref. \cite{Langmuir:1959}. In this work, the ions were cooled by means of a light background gas and by varying the trap potential and the cooling force, repeated crystallizations and meltings could be monitored. Phase transitions have been observed for laser-cooled ${\rm Mg^+}$ as well as ${\rm Hg^+}$ ions in a Paul trap in Refs. \cite{Walther:1987}, and \cite{Manney:1987}, respectively. The phase transition of a two ion crystal to a cloud has been explained by the nonlinear aspects of the Paul trap, i.e. the melting of a crystal has been interpreted as an order to chaos transition induced by ion-ion collisions \cite{Walther:1988,Brewer:1988,Walther:1989,Henshaw:1990}. Further research on nonlinear phenomena in trapped ion systems dealt with transient chaotic regimes preceding the transition point of the Paul trap \cite{DeVoe:1990}, frequency-locked motion of two ions  in the Paul trap \cite{Brewer:1993} and with nonlinear effects in the dynamic Kingdon trap \cite{Blümel:1995}. In Ref. \cite{Fishman:2004a,Fishman:2004b,Fishman:2006} the dynamics of Coulomb chains with many ions in a harmonic potential has been studied. The authors evaluate analytically the long-wavelength modes and the density of states in the short-wavelength limit. Using these results they are able to derive the critical transverse frequency of the trap below which the chain is unstable with respect to excitations of transverse vibrations.

In this work we study the escape dynamics of a chain of ions in a finite box-like smooth potential. The non-equilibrium properties of this one-dimensional system of long-range interacting ions will be explored for the case when the trapping potential is instantaneously reduced (we will denote this in the following by a ``ramp'' of the potential well) such that the emission of particles from the trap is possible. Our aim is to unravel and analyze the non-equilibrium dynamics as well as the escape properties of ions from such a one-dimensional potential well. We find that above an energetic threshold the energy shell reaches out to infinity thus making the escape of ions from the potential well possible. The emission process is dominated by the unstable periodic orbits and their manifolds in the sense that they lead to dominant peaks in the distribution of the kinetic energy of the escaping ions. Ramping up the potential we observe a strong correlation between the escape time and the kinetic energy of the emitted ions. Most importantly we show that the ionic emission occurs in a sequence of synchronized pairs whose properties are tunable via the trap parameters. In particular by a ramp of the well to a corresponding depth the number of emitted pairs as well as the time difference between consecutively emitted pairs and their kinetic energies can be adjusted. Accordingly, this escape process from the trap yields a deterministic source of correlated ion pairs with selectable dynamical properties. Such a source could be used for performing scattering experiments in which one needs energetically and spatially correlated particles.

This paper is organized as follows: In Sec. \ref{sec:setup} we introduce the Hamiltonian and describe the corresponding scaling transformations. Sec. \ref{sec:fewions} is devoted to the study of the classical phase space of a single and two ions in the static potential well. In Sec. \ref{sec:ramping} the escape dynamics of many ions from the trap after a ramp of the potential depth is discussed. Finally, our summary is provided in Sec. \ref{sec:summary}.

\section{Setup}\label{sec:setup}
\begin{figure}
\includegraphics[width=\columnwidth]{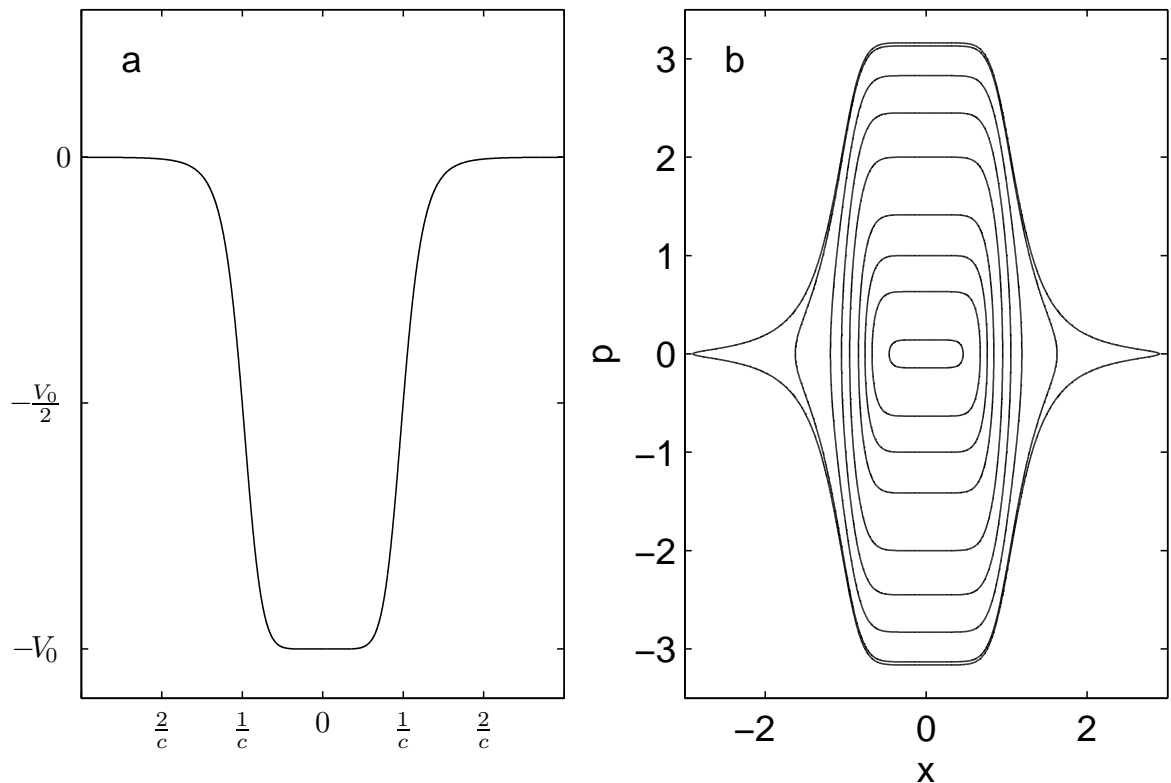}
\caption{\label{fig:fig1}  (a) Potential $V(x)$ as given by Eq. \eref{eq:potential}. (b) Phase space portrait for bounded dynamics of a single ion in $V(x)$.}
\end{figure}

The Hamiltonian (in Gaussian units) describing the dynamics of $n$ Coulomb-interacting particles with equal charge $q$ and mass $m$ in an external potential $V(x)$ is given by
\begin{equation}\label{eq:hamil}
 H(\vec{x},\vec{p})=\sum\limits_{i=1}^{n}\frac{p_{i}^{2}}{2m}+\sum\limits_{i=1}^{n} V(x_i)+\frac{1}{2}\sum\limits_{i=1}^{n}\sum\limits_{j\neq i}\frac{q^{2}}{|x_{i}-x_{j}|},
\end{equation}
where $\vec{x}=(x_1,\dots,x_n)$, $\vec{p}=(p_1,\dots,p_n)$ are the positions and the momenta of the ions, respectively. The ions are numbered according to their alignment, i.e. $x_1$ is the position of the leftmost ion and $p_1$ its momentum etc. For $V(x)$ we choose the following functional form
\begin{equation}\label{eq:potential}
 V(x)=-\frac{V_{0}}{1+\left(cx\right)^8},
\end{equation}
which corresponds to a smooth potential well of depth $V_0$. $c$ fixes the width of the well via $V(1/c)=-V_0/2$. Fig. \ref{fig:fig1} (a) shows the potential $V(x)$. For this functional form of the potential the Coulomb crystal configuration is an almost equidistantly spaced ion chain. The one-dimensional confinement of the dynamics can be achieved for example in RF-traps by appropriately adjusting the trap frequencies along the transverse dimensions, i.e. when the kinetic energy of all ions is small compared to the transversal harmonic confinement frequency \cite{Fishman:2004a,Fishman:2004b,Fishman:2006}. Hence, the dynamics of the system is one-dimensional. Apparently, the parameter space of the Hamiltonian is four-dimensional: the depth of the well $V_{0}$, the width of the well determined by $\frac{1}{c}$, the mass of the ions $m$ and the charge of the ions $q$.
It is therefore useful to apply a scaling transformation to dimensionless variables:
\begin{equation*}
\quad x \rightarrow x^\prime =cx, \quad p \rightarrow p^\prime =\frac{c}{m}p
\end{equation*}
If the Hamiltonian is scaled additionally according to $H \rightarrow H^\prime =\frac{c^2}{m}H$ the result reads
\begin{equation}
H^\prime(\vec{x}^\prime ,\vec{p}^\prime)=\sum\limits_{i=1}^{n}\frac{{p^\prime_i}^2}{2}-\sum\limits_{i=1}^{n}\frac{V_0^{\prime}}{1+{x^\prime _{i}}^8}+\frac{1}{2}\sum\limits_{i=1}^{n}\sum\limits_{j\neq i}\frac{{q^\prime}^{2}}{|x^\prime _{i}-x^\prime _{j}|},
\end{equation}
with a scaled charge $q^\prime =\sqrt{\frac{c^3}{m}}q$ and potential $V_0^\prime =\frac{c^2}{m} V_0$. Obviously, this transformation yields a reduction of the dimension of the parameter space to two: the potential depth $V^\prime _0$ and the charge $ q^\prime$. For reasons of clarity the prime is omitted in the following. The corresponding Hamiltonian equations of motion
\begin{equation}
 \dot{x_i}=\frac{\partial H}{\partial p_i}, \quad \dot{p_i}=-\frac{\partial H}{\partial x_i}
\end{equation}
are solved numerically using the Burlisch-Stoer algorithm \cite{NRC:1992}.

\section{Phase space for few ion systems}\label{sec:fewions}
\begin{figure}
\includegraphics[width=\columnwidth]{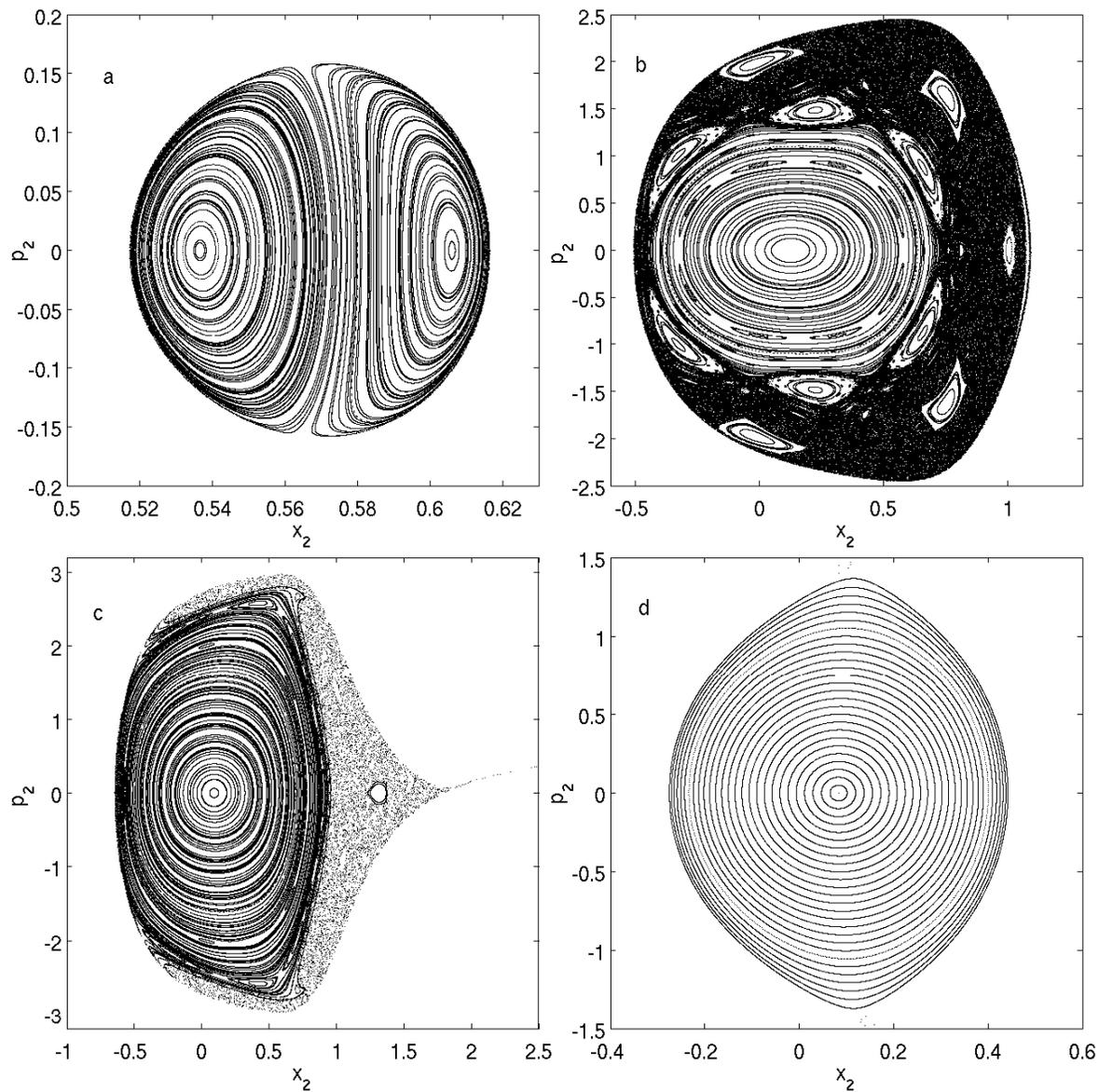}
\caption{\label{fig:fig2} PSS for two ions in the static trap for (a) $E=-9$, (b) $E=-6$, (c) $E=-4.6$ and (d) $E=-4$.}
\end{figure}
\begin{figure}
\includegraphics[width=\columnwidth]{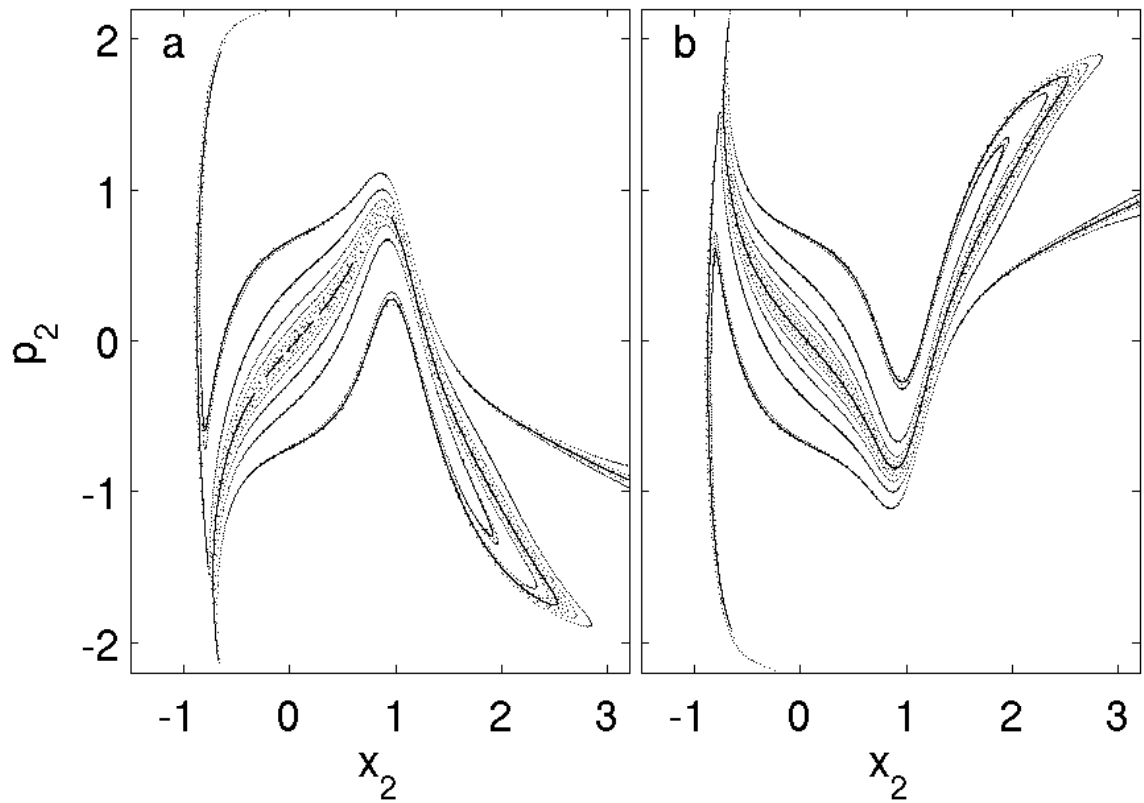}
\caption{\label{fig:fig3} Flow of the stable (a) and unstable (b) manifold belonging to the period one unstable fixed point at $U=(x_2=0.0625,p_2=0)$.}
\end{figure}

Let us now turn to the discussion of the phase spaces for setups with few ions in the potential well. All results in this section are obtained for $q=1$ and $V_0=5$. Since the energy is a constant of the motion the one ion system is integrable and the two dimensional phase space can be visualized straightforwardly. Fig. \ref{fig:fig1} (b) shows the level curves for bounded oscillatory motion which are simply the contour lines of the energy. Obviously, the initial momenta for these trajectories must obey $p<\sqrt{2V_0}$. For two ions in the trap a representative visualization is still possible. Due to energy conservation the dynamics in the four dimensional phase space is confined to a three-dimensional energy shell. The lowest energy state of two ions in the static well is the so-called {\it Coulomb crystal} and corresponds to a single point in phase space. For our choice of the parameters for $V_0$ and $q$ this configuration possesses the energy $E_C=-9.0126$. The energy shells in phase space with $E>E_C$ are visualized by the corresponding Poincar\'{e} surfaces of section (PSS) using the condition $p_1=0$. The PSS is made such that we report crossings from $p_1>0$ to $p_1<0$, i.e. we plot the momentum against the position of the right (second) ion, when the left one reaches its inner turning point. Fig. \ref{fig:fig2} shows the PSS for different values of the total energy. For $E=-9$ the dynamics is completely integrable (Fig. \ref{fig:fig2} (a)). At $S_1=\left(x_2=0.537,p_2=0\right)$ and $S_2=\left(x_2=0.606,p_2=0\right)$ there are stable period one orbits, which correspond to the two {\it normal modes}, i.e. in the case of $S_1$ both ions oscillate with opposite phase such that the center of mass is at rest, whereas $S_2$ belongs to the {\it center-of-mass} mode (oscillation in phase). With increasing energy the dynamics becomes richer. Elliptic islands embedded in a chaotic sea appear (Fig. \ref{fig:fig2} (b)), whose area in the PSS grows until $E=-5$. For even larger energies one observes a growth of the central integrable part. For $E=-4.59$ initial conditions in the chaotic sea are able to leave the well, i.e. the energy shell reaches in the $x_1, \, x_2$-direction to infinity which can be seen as an excrescence on the right hand side in Fig. \ref{fig:fig2} (c). For higher initial energies the size of the opening of the phase space towards infinity gets bigger which allows for a quicker escape of the particles from the chaotic sea to infinity, e.g. in Fig. \ref{fig:fig2} (d) the initial conditions outside of the shown area of closed curves belong to trajectories where the ions leave the trap rapidly. Furthermore, with increasing energy the remaining integrable part shrinks, until for $E\geq-3.6$ no integrable structure in phase space is present. In the energy range $E\geq-3.6$ unstable periodic orbits and their manifolds organize the dynamics in phase space. In order to detect them we apply a two-dimensional Newton-Raphson-method \cite{NRC:1992} in the PSS. Fig. \ref{fig:fig3} shows the stable and unstable manifolds of an unstable period one orbit located at $U=(x_2=0.063, p_2=0)$ for $E=-2$. The flow of the unstable manifolds is constructed by distributing a whole line segment of initial conditions along the directions of the eigenvectors of the underlying monodromy matrix. Both manifolds reach in the $x$-direction out to infinity. In Fig. \ref{fig:fig3} this is only partially visible because the part of the manifolds which reach out to infinity is cut off in the $x$-range of the well. Below we will see that the manifolds are important for the understanding of the escape of ions from the well.

\subsection{Escape of a single ion from the well}\label{sec:escape}
In this section the properties of the single ion escape process from the well are discussed, i.e. one ion possesses via the Coulomb interaction with its partner sufficient energy such that it can be emitted. We do not consider the two ion escape processes which is possible for $E\geq 0.42$ because in this high-energy regime the immediate escape of both ions dominates and no new time scales or processes emerge. In order to avoid artificial transient effects we have to ensure that the initial conditions are distributed uniformly over the energy shell. To this end, we briefly describe in the following the preparation of the initial ensemble. First, the positions of the ions $(x_1,x_2)$ are distributed randomly in the interior of the well, which we define to be $|x|<2.5$. Due to energy conservation, the corresponding momenta lie on a circle with radius $r(x_1,x_2)=\sqrt{p_1^2+p_2^2}=\sqrt{2m(E-V(x_1,x_2))}$ for each pair $(x_1,x_2)$. On these circles in momentum space the points are distributed with the same density as in position space. Altogether, this procedure yields a uniform distribution of quadruples $(x_1,x_2,p_1,p_2)$ on a given energy shell. From the previous discussion of the phase space properties it is known that unbounded motion is only possible for energies $E\geq -4.59$. Accordingly, we distribute approximately $10^5$ initial conditions uniformly over different energy shells with $E\geq-4.59$ and record the escape time and the kinetic energy at the border of the well (note that all observed effects and phenomena are qualitatively not affected by the position of the detector, where the escape time and kinetic energy are recorded). Generally, with increasing initial energy the region in phase space allowing for escape grows and the particles leave the well more quickly. Nevertheless, we observe also events where it takes long time intervals until emission occurs.
\begin{figure}
\includegraphics[width=\columnwidth]{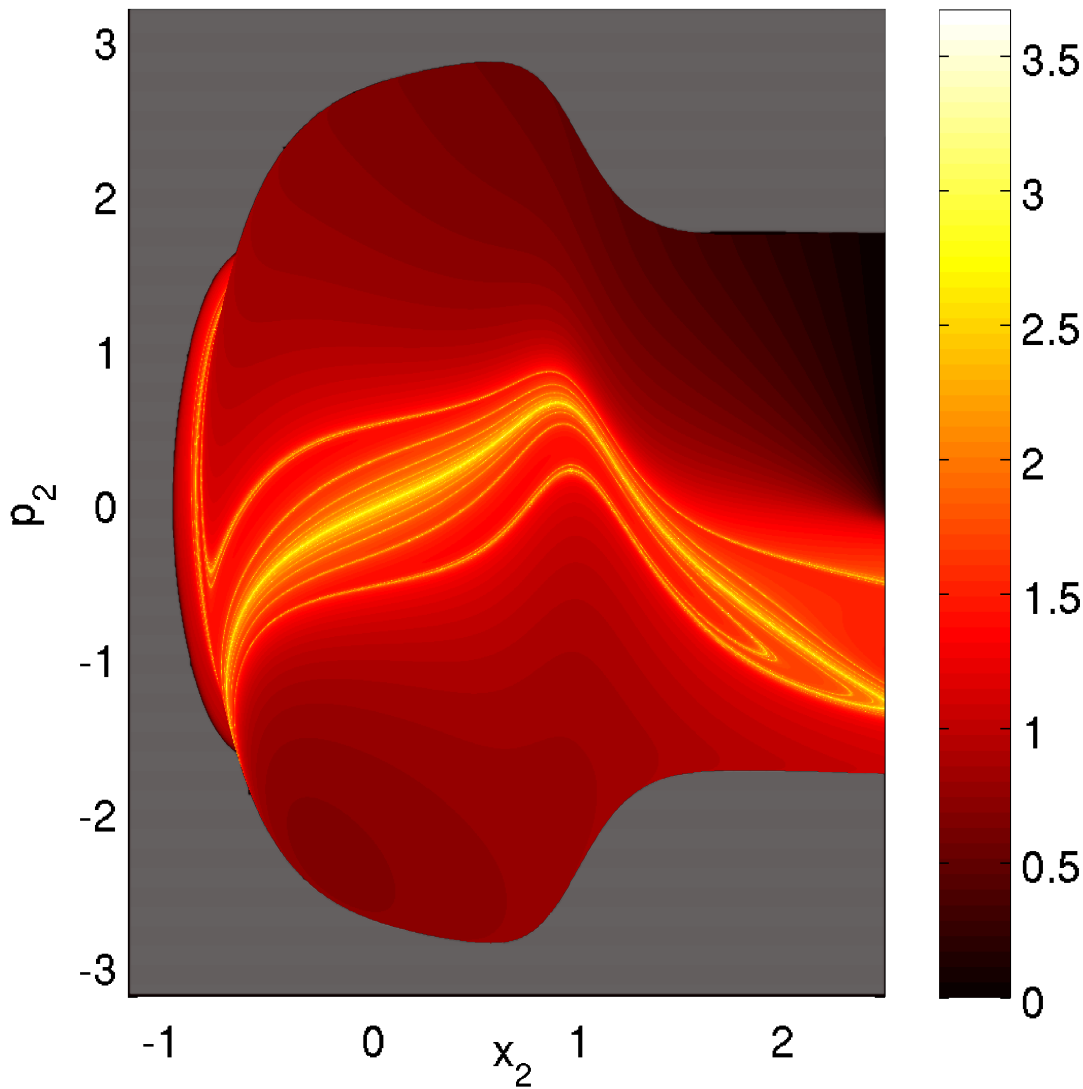}
\caption{\label{fig:fig4} Logarithm of the escape time $\log t$ for initial conditions in the PSS $p_1=0$ on the energy shell $E=-2$.}
\end{figure}
Fig. \ref{fig:fig4} shows the escape times on a logarithmic scale for initial conditions on the energy shell $E=-2$ which lie additionally in the plane $p_1=0$ formed by the PSS. The border of the grey area defines the border of the energy shell. A comparison of Fig. \ref{fig:fig4} with Fig. \ref{fig:fig3} (a) shows that the longest escape times belong to initial conditions close to the stable manifold of the period one orbit $U$. This orbit is marginally unstable because it originates from $S_1$ which is one of the dominant stable periodic orbits. Consequently, the particles starting in the vicinity of the stable manifold of $U$ approach the fixed point, stay in its vicinity for a long time and escape afterward via the unstable manifold leading to infinity. Fig. \ref{fig:fig5} (a) shows the probability density of the kinetic energy of the escaping ion $\rho(E_{\rm{kin}})$ for initial energy $E=-4$, where $E_{\rm{kin}}$ is recorded at the border of the well $|x|=2.5$. $\rho(E_{\rm{kin}})$ possesses a sharp peak at the maximal value for the kinetic energy, which is determined by energy conservation. Obviously, the Coulomb term as well as the potential energy of the escaped ion is zero. Furthermore, the minimum energy of the residual ion is $E_r=-V_0$. Thus the maximum kinetic energy of the escaped ion is $E_{\rm{kin,max}}=E+V_0$ which amounts in our case to $E_{\rm{kin,max}}=1$. Since there is still a remaining Coulomb contribution (the detector is at a finite distance from the trap) for the distribution shown in Fig. \ref{fig:fig5} (a), the maximum kinetic energy $E_{\rm{kin,max}}$ is smaller, i.e. $\rho(E_{\rm{kin}})$ has a sharp cut-off at $E_{\rm{kin}}=0.65$ and not at $E_{\rm{kin}}=1.0$. The general appearance of $\rho(E_{\rm{kin}})$, namely an increase followed by a plateau-like behaviour, can be understood intuitively. For $E=-4$ the phase space contains a regular island of bounded dynamics in the potential well (see Fig. \ref{fig:fig2} (d)). Consequently, the ions starting in this integrable structure do not contribute to the escape. Due to the fact that these initial conditions correspond to small kinetic energies, the distribution $\rho(E_{\rm{kin}})$ decreases for $E_{\rm{kin}} \rightarrow 0$.
\begin{figure}
\includegraphics[width=\columnwidth]{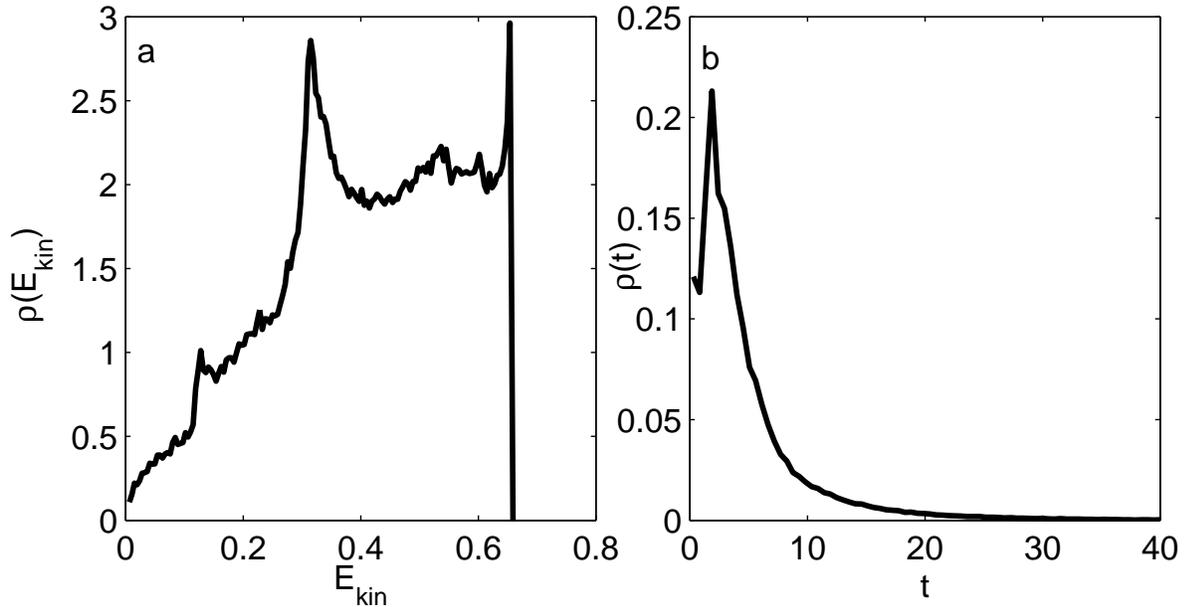}
\caption{\label{fig:fig5} Probability distribution of kinetic energies (a) and escape times (b) of the escaping ion for the energy shell $E=-4$ and $V_0=5$.}
\end{figure}
Fig. \ref{fig:fig5} (b) shows the distribution of escape times $\rho(t)$ which has a dominant peak followed by a steep drop. With decreasing initial energy the tail of the distribution becomes more pronounced until a collapse occurs for $E=-4.6$ and no escape of ions is possible anymore. The peak of $\rho(t)$ corresponds to the immediate emission process. A further analysis of the correlation between the kinetic energy (Fig. \ref{fig:fig5} (a)) and the escape time (Fig. \ref{fig:fig5} (b)) reveals that the peaks in $\rho(E_{\rm{kin}})$ belong to initial conditions with particularly long escape times and thus to particles starting in the vicinity of the stable manifolds of marginally unstable fixed points (see Fig. \ref{fig:fig4} and the corresponding discussion of the marginally unstable periodic orbits). The width of these peaks of $\rho(E_{kin})$ corresponds to the volume in phase space which gets transported along the unstable manifold out of the trap. The structure of these manifolds becomes more complicated with decreasing energy leading to a sequence of peaks in $\rho(E_{\rm{kin}})$. Finally, the channel of the energy shell towards infinity becomes thinner. Accordingly, the distribution of kinetic energies develops into a delta distribution located at $E_{\rm{kin}}=0.4$ for $E\rightarrow -4.6$.

\section{Escape dynamics after a ramp of the well}\label{sec:ramping}
In this central section of our work the dynamics following a ramp of the well with many ions is explored. A ramp means that the depth of the well is reduced instantaneously. Due to the fact that the number of degrees of freedom increases according to the number of particles, the energy shells of the static system for many ions are topologically complicated objects of high dimensionality. Hence a uniform distribution of initial conditions on an energy shell is not only difficult to achieve but in particular also not a natural starting point for an experimental setup. Instead we use as initial conditions an ensemble which is sampled in position space randomly in small intervals $\Delta x$ around the Coulomb crystal configuration with all initial momenta set to zero. For the interval widths $\Delta x$ we choose, as a typical value, the twentieth part of the mean crystal distance. Instead of the uniform distribution on a given energy shell this is indeed a more realistic scheme with respect to an experimental preparation of such a setup. For the charge of the ions $q$ and the depth $V_0$ we choose $q=1$ and $V_0=30$ such that the Coulomb crystal exists for up to ten ions in the well. In a second step the well is ramped up by $\Delta V$ and the resulting dynamics is observed and analyzed.

\paragraph{Three ions}
For $\Delta V < 20$ the ions oscillate with comparatively small amplitudes and no significant energy transfer among them occurs. However, for a further reduction of the well's depth $V_0$ the Coulomb interaction gains importance such that the dynamics becomes more and more chaotic: for a larger $\Delta V$ the ions oscillate with larger amplitudes and thus the relative distance between them can become small. Therefore, the probability increases that in the course of the dynamics after the ramp an ion gains enough kinetic energy to be emitted. Still, due to energy conservation there is an energy threshold for a single ion-escape to become possible. Nevertheless, above this threshold not every initial condition leads to an emission-event. For $V_0=2.3$ the escape of a single ion becomes energetically possible. For this energy which is slightly above the mentioned threshold value the escape process is typically characterized by a phase of motion during which two neighbouring ions move in the same direction. This leads to a strong enhancement of the kinetic energy of the third outermost ion which gets ``kicked'' out of the potential well.
After the escape of one ion the relevance of the Coulomb repulsion decreases and the remaining ions start quasi-periodic oscillations around the equilibrium positions of the two-ion Coulomb crystal.
\begin{figure}
\includegraphics[width=\columnwidth]{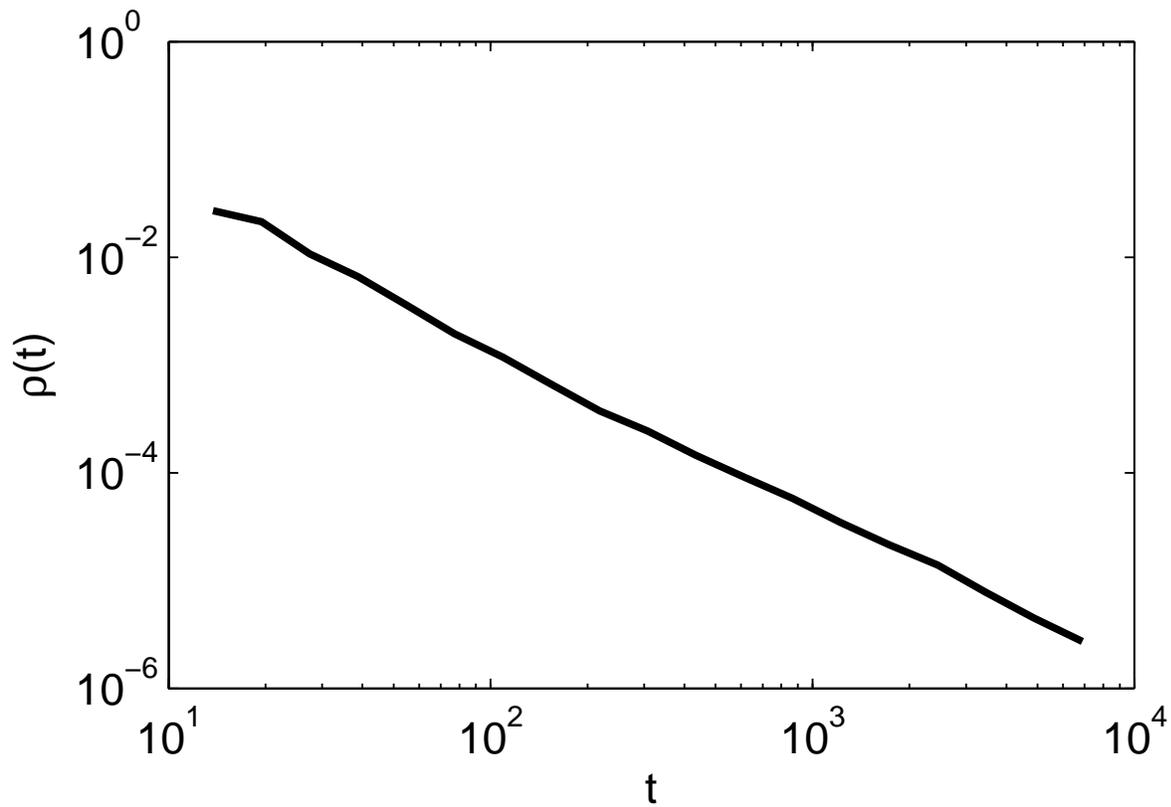}
\caption{\label{fig:fig6} Probability distribution of escape times $\rho(t)$ of a single ion when three ions are loaded into the trap and the well is ramped from $V_0=30$ to $V_0 =2.1$.}
\end{figure}
In Fig. \ref{fig:fig6} the distribution of the escape times for a ramp to $V_0=2.1$ is presented, which shows asymptotically (large $t$) an algebraic decay. The long tail is due to initial conditions with total energies hardly above the energy threshold. Consequently, for these events the energy exchange process leading to an ion-escape has to be very efficient meaning that two ions must transfer a large fraction of their kinetic energy to the third ion such that the latter can leave the well. Accordingly, a larger $\Delta V$ of the ramp of the potential does not necessarily yield a higher kinetic energy of an individual escaping ion because the emission occurs also in case a smaller portion of the kinetic energy of two ions is transfered to the third one. By a ramp of the well to even smaller depths $V_0\leq 1.5$ the two ion emission process becomes possible. Finally, for $V_0 \leq 1.3$ all initial conditions yield two ion escape. In this regime the emission occurs generally in a synchronized manner, i.e. as pairs where the difference between the escape times $\Delta t_p =t_1-t_2$ of the two ions of a pair is small compared to all other time scales (see below for a more detailed discussion of this process). We remark that by fine-tuning $\Delta V$ it is also possible to achieve the situation where $\Delta t_p$ can become large. Such a two-ion emission occurs if the ramping of the well $\Delta V$ is tuned to be only slightly above the energy threshold for two-ion emission. These events are rare and their fraction remains below 0.75 \% in the regime $1.5 \geq V_0 \geq 1.3$.

\begin{figure}
\includegraphics[width=\columnwidth]{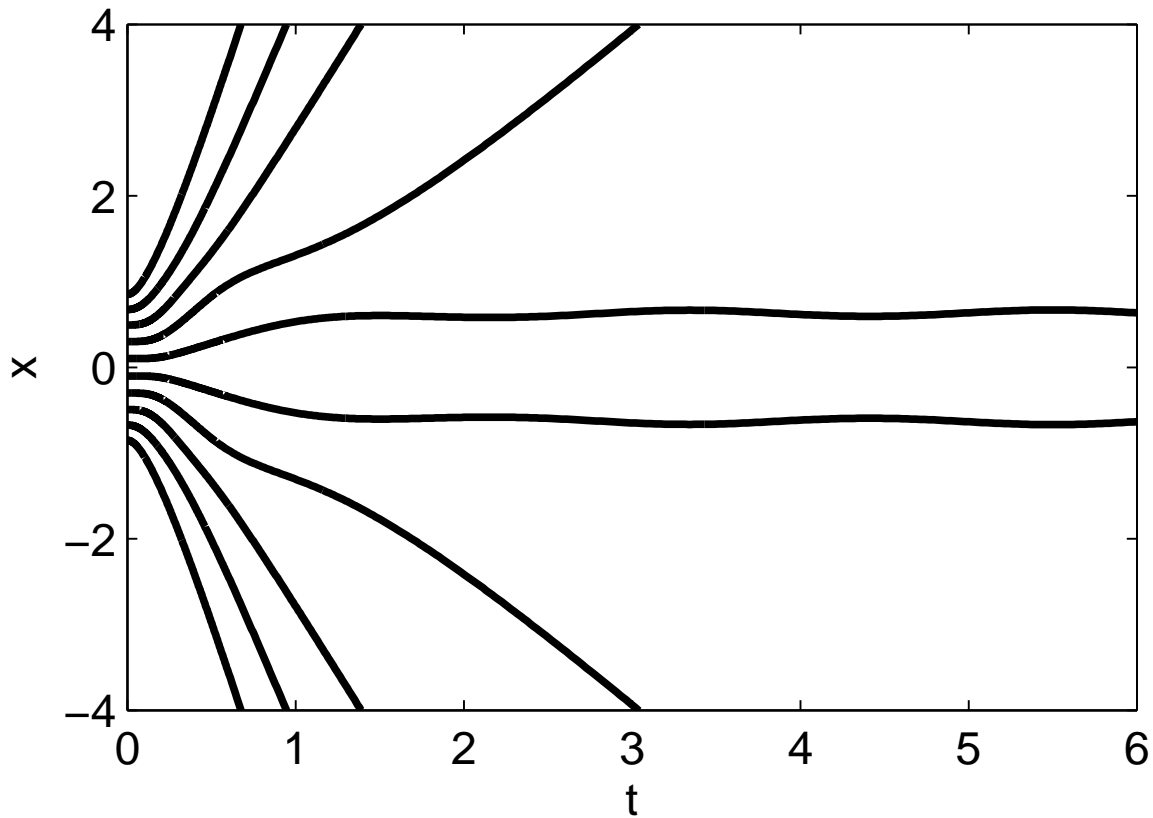}
\caption{\label{fig:fig7} Trajectories of the ten ions when the potential depth is instantaneously reduced to $V_0=2$. For $V_0=30$ the initial condition corresponds to the Coulomb-crystal.}
\end{figure}

\paragraph{Ten ions}
For ten ions the Coulomb interaction is stronger and thus we expect a larger impact of the instantaneous reduction of the potential depth. Fig. \ref{fig:fig7} shows the dynamics of trajectories for a ramp of the trap from $V_0=30$ to $V_0=2$. The single initial condition used for the simulation corresponds to the configuration of the Coulomb crystal for $V_0=30$. In this case we observe that eight ions leave the well. Obviously, the emission of the ions happens in a sequence of four pairs. Every time a pair is emitted the corresponding two ions leave the trap in opposite directions.

\begin{figure}
\includegraphics[width=\columnwidth]{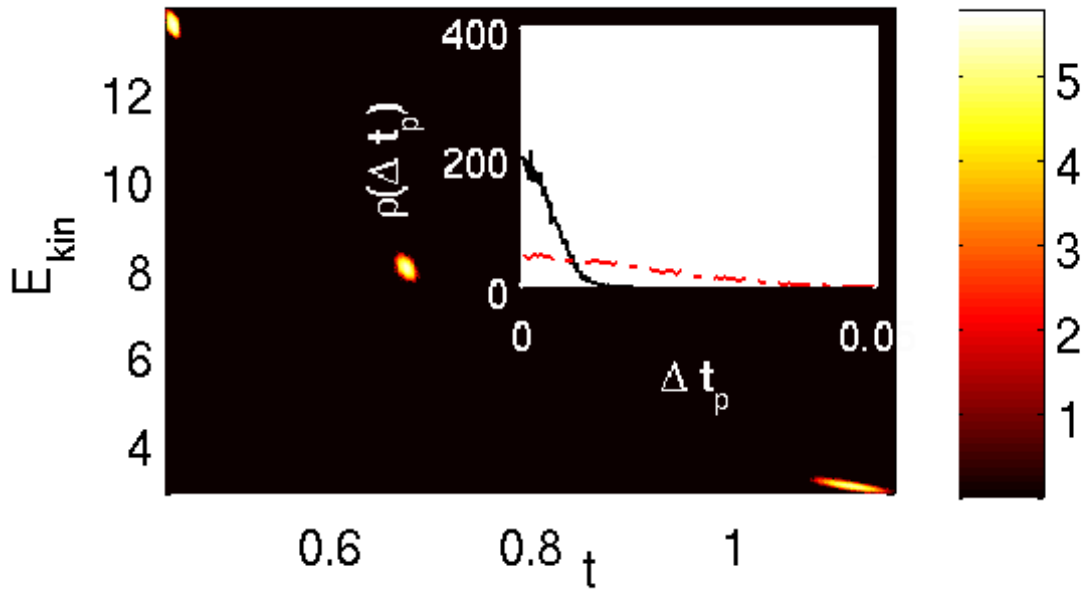}
\caption{\label{fig:fig8} (Color online) Logarithm of the frequency distribution of the kinetic energy and the escape time $\log(\rho(E_{\rm{kin}},t))$ for a ramp from $V_0=30$ to $V_0=4$, where six ions leave the well. Each white spot belongs to two emitted ions. In the inset the distribution of the differences of the escape times $\rho(\Delta t_p)$ between the first and the second ion (solid line) and the fifth and the sixth ion (dashed line) are shown.}
\end{figure}

In the following we study the robustness of this behaviour with respect to the choice of the initial ensemble, i.e. for a random sampling in the vicinity of the Coulomb crystal. Accordingly, we take $10^4$ different initial ensembles of ions, ramp up the trap by $\Delta V$ and record the kinetic energy as well as the escape time of the emitted ions. Thus all following results are ensemble properties. Fig. \ref{fig:fig8} shows the logarithm of the frequency distribution $\log(\rho(E_{\rm{kin}},t))$ of the kinetic energy and the escape time for a ramp from $V_0=30$ to $V_0=4$. In this case six ions escape from the trap. Each of the three bright dots in Fig. \ref{fig:fig8} corresponds to two emitted ions. In the inset the frequency distribution of the time differences $\rho(\Delta t_p)$ between the escape of the first and the second ion and the fifth and the sixth ion is shown. $\rho(\Delta t_p)$ belonging to the third and fourth ion is almost identical to the corresponding distribution for the first and the second ion. Obviously, $\Delta t_p$ is very small compared to the interpair emission time intervals $\Delta t_{pp}$. Additionally, the distribution $\rho(\Delta t_p)$ decreases rapidly, i.e. to a very good approximation the emission process can indeed be regarded as pairwise. An intuitive physical explanation of this result is given in the following: Before the ramp the initial ensembles are spatially close to the highly symmetric Coulomb crystal configuration. Due to the instantaneous reduction of the potential depth much potential energy is transfered into the system. Subsequently, these now excited chains relax by the emission of ions. Since the symmetry of the trap is preserved by the ramp, the forces acting on the outermost ions immediately after the instantaneous reduction of the potential have the same absolute value but opposite signs and consequently the ions escape in opposite directions within a small time-interval. Furthermore, Fig. \ref{fig:fig8} reveals that $\Delta t_{pp}$ increases within a sequence of pairs whereas the kinetic energy decreases. This behaviour can be explained as follows. The force which pushes the ions outwards is largest for the first pair and decreases with every subsequently emitted pair. Additionally, every escaping pair is slowed down by the Coulomb repulsion of the previously emitted ions. In Fig. \ref{fig:fig7} these two characteristics can be seen as a ``fanning out'' of the trajectories with a decreasing slope. Thus the escape velocity becomes smaller with every emitted pair. By ramping the trap to different potential depths in the regime where the emission of six ions occurs, the time difference between the consecutive pairs $\Delta t_{pp}$  can be tuned, i.e. $\Delta t_{pp}$ decreases with increasing $\Delta V$. Moreover, since the total energy of the ensemble after the ramp is higher for larger $\Delta V$, the kinetic energy of each emitted pair increases as well.

So far we have discussed the situation when the ramp leads to an emission of an even number of ions. Yet, it is also possible that after the ramp only an odd number of ions can leave the trap. For these cases the scenario is the following. First one observes the emission of a maximal even number of ions in terms of synchronized pairs according to the discussion provided above. Afterwards, a single ion leaves the well. Its escape from the trap can be delayed significantly with respect to the emission of the last pair. Such an event occurs when the ramp is chosen only slightly above the necessary energy threshold and can be explained straightforwardly. After the emission of the pairs the remaining ions oscillate chaotically in the trap. Accordingly, it can take a very long time until, in the course of the dynamics, the randomly distributed kinetic energy accumulates onto a single ion which can then be emitted. Consequently, the distribution of the kinetic energies of the last single escaping ion is not peaked at a certain value as for the pairwise emission but it shows an exponentially decreasing tail.

\begin{figure}
\includegraphics[width=\columnwidth]{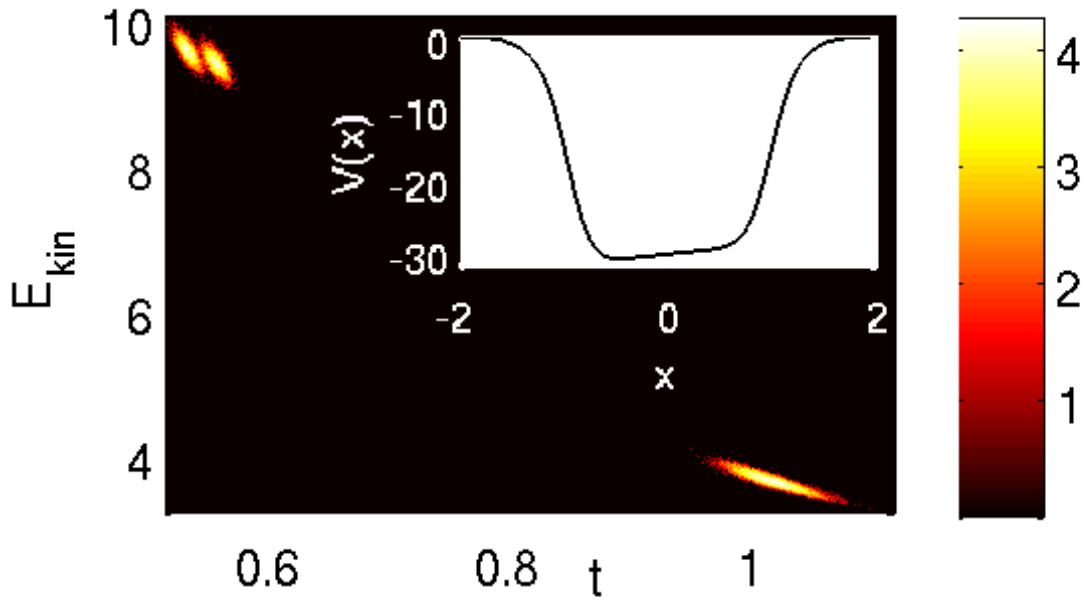}
\caption{\label{fig:fig9} (Color online) Logarithm of the frequency distribution of the kinetic energy and the escape time $\log(\rho(E_{\rm{kin}},t))$ for a ramp from $V_0=30$ to $V_0=4$ of an asymmetric well. In this case four ions are emitted from the trap. The inset shows the potential described by Eq. \eref{eq:pot_asym} for $V_0=30$ and $\alpha=\frac{1}{20}$}
\end{figure}
Let us now discuss the symmetry robustness of the synchronized pair emission process. To this end we modify the functional form of the potential by multiplying it with an asymmetric exponential term, i.e.
\begin{equation}\label{eq:pot_asym}
V(x)=-\frac{V_0 \, e^{-\alpha x}}{1+x^8}.
\end{equation}
In Fig. \ref{fig:fig9} the modified potential Eq. \eref{eq:pot_asym} is shown in the inset for $V_0=30$ and $\alpha=\frac{1}{20}$. For $x\rightarrow -\infty$ we have $V(x) \rightarrow -\infty$. However on the scale on which the dynamics is observed this behaviour is suppressed by the large power of the denominator and the fact of a small value of the prefactor $\alpha$. Due to the parity asymmetry of the potential the Coulomb crystal and thus the sampling of the initial ensemble is also asymmetric with respect to $x=0$. Fig. \ref{fig:fig9} shows $\log(\rho(E_{\rm{kin}},t))$ for a ramp from $V_0=30$ to $V_0=9$. In this case four ions escape from the trap. The first ion leaves the well in positive $x$-direction at $t\approx0.52$ with kinetic energy $E_{\rm{kin}}\approx9.59$ (see Fig. \ref{fig:fig9}). The second ion which has less kinetic energy $E_{\rm{kin}}\approx9.22$ escapes to $-\infty$ at $t\approx0.56$. Afterwards, the emission of a pair with $E_{\rm{kin}}\approx3.68$ occurs at $t\approx1.02$. This behaviour can be traced back to the shape of the trap. Obviously, due to the asymmetry of the potential the leftmost ion has a lower potential energy. Thus compared to the rightmost ion a larger fraction of its kinetic energy is needed for the emission from the trap after the ramp of the potential depth. Since the potential difference due to the asymmetry of the trap decreases towards the origin, the difference of potential energies is less for the next two outer ions. Thus they escape pairwise again. For a larger ramping $\Delta V$ one observes also an energetic and temporal separation of the emission for the later escape events, that is for an appropriate choice of $\Delta V$ the third and fourth ion etc can be separated as well. Due to the fact that these results are obtained for a random sampling of the ions' initial position in the vicinity of the corresponding Coulomb crystal we expect that effects like micromotion do not alter qualitatively the results. Furthermore, this statement is supported by our numerical observation that no significant changes occur when the ions are additionally given a small initial momentum, i.e. the emission properties of the ions from the trap are very robust.

\section{Summary}\label{sec:summary}
We have studied the nonlinear escape dynamics of long-range Coulomb interacting ions from a one-dimensional trap. As a necessary prerequisite we have first accounted for the few ion setup, its phase space structure, and observed that the phase space of two ions in the static potential well can be regular, mixed or chaotic depending on the energy. For energies larger than a certain threshold the energy shell extends towards infinity and consequently the escape of ions from the trap is possible. It has been elucidated that the manifolds of unstable periodic orbits play a crucial role in this emission process. Particles starting in the vicinity of the stable manifold show the longest escape times from the well and contribute to dominant peaks in the distribution of the kinetic energy of the escaping ions.

Our main result concerns the escape dynamics of an  ensemble of ions initially in a configuration close to the Coulomb crystal when the potential depth is decreased by an instantaneous ramp of the trap. We have observed that the emission of an even number of ions occurs in a sequence of pairs. Accordingly, every time a pair is emitted the corresponding two ions escape in opposite directions. A deeper analysis of the escape time and the kinetic energies reveals a strong correlation between both quantities, that is each pair leaves the trap with a definite kinetic energy. By adjusting the value of the ramp $\Delta V$ the number of the emitted pairs, the time differences between the consecutive pair emission events and the average kinetic energy of each pair can be varied. If the ramp is chosen such that an odd number of ions can escape, one observes after a maximal even number of ions escaping pairwise the emission of a single ion. Its escape can be delayed significantly. Moreover, the kinetic energy distribution possesses an exponentially decreasing tail, because the dynamics leading to its escape is chaotic. Furthermore, we have studied the symmetry robustness of the escape properties. It has been found that parity asymmetry of the potential leads to a temporal and energetic separation of the pairwise emission. We have also considered the robustness of our results with respect to variations of the potential by choosing a less steep and box-like potential $V(x)$ and found that this does not alter qualitatively the results.

Obviously, the ramp of the trap from the regime where the initial ensemble is confined in a configuration close to the Coulomb crystal to a lower potential depth can be exploited to create a temporally synchronized source of ions. By appropriately choosing $\Delta V$ a desired number of emitted ions (pairs) and their escape properties like escape time and kinetic energy can be adjusted.

\ack
The authors acknowledge gratefully helpful discussions with M. Drewsen and F. K. Diakonos. This work has been performed within the Excellence Cluster Frontiers in Quantum Photon science, which is supported by the Joachim Herz Stiftung. Financial support by the DAAD in the framework of an exchange program with Greece is acknowledged. Part of this work was financially supported in the framework of the excellence initiative of the federal and state government of Germany, i.e., the FRONTIER program of the university of Heidelberg.

\end{document}